\documentclass[aps,prb,reprint,twocolumn]{revtex4-1}

\usepackage[english]{babel}
\usepackage{amsmath,amsfonts,amsthm,bm,braket}
\usepackage[letterpaper,top=2cm,bottom=2cm,left=3cm,right=3cm,marginparwidth=1.75cm]{geometry}

\usepackage{graphicx}
\usepackage[colorlinks=true, allcolors=blue]{hyperref}
\usepackage{xurl}
\hypersetup{breaklinks=true}
\usepackage{float}
\usepackage{physics}
\usepackage{tikz}
\usetikzlibrary{trees}
\usepackage{caption}
\usepackage{subcaption}
\usepackage{verbatim}
\usepackage{bbold}
\usepackage{comment}
\usepackage{adjustbox}

\begin{document}

\title{Quantum advantage beyond entanglement in Bayesian game theory}
\author{A. Lowe}
\email{lowea3@aston.ac.uk}
\affiliation{Aston University, School of Informatics \& Digital Engineering, Birmingham B4 7ET, UK}

\date{\today}

\begin{abstract}
Quantum discord has been utilised in order to find quantum advantage in an extension of the Clauser, Horne, Shimony, and Holt (CHSH) game. By writing the game explicitly as a Bayesian game, the resulting game is modified such the payoff's are different, and crucially restrictions are imposed on the measurements that Alice and Bob can perform. By imposing these restrictions, it is found that there exists quantum advantage beyond entanglement for a given quantum state. This is shown by decomposing the expected payoff into a classical and quantum term. Optimising over the expected payoff, results in the classical limit being surpassed. This gives an operational framework in order to witness and determine quantum discord.
\end{abstract}

\maketitle

\section{Introduction}
The search for quantum advantage has spanned a range of different fields, with particular focus on finding systems where non-locality offers significant quantum advantage \cite{quantumadvantage,doi:10.1126/science.abe8770,bellbayes}. However, there has been little effort in attempting to use other forms of quantum correlations in order to witness quantum advantage. In particular, quantum discorded states remain under utilised in the literature, both in quantum game theory and quantum networks. Quantum discord was proposed in tandem by Ollivier and Zurek \cite{discord} and Henderson and Vedral \cite{discord1}. Quantum discord proposes that quantum correlations can exist even when there is no entanglement, which is often the case for mixed separable states. An additional characteristic of quantum discord is that it is relatively robust against perturbations \cite{PhysRevA.80.024103} compared to entanglement and non-locality, which means states which are discorded are more experimentally feasible.

A significant drawback is that quantum discord is difficult to compute analytically, so various methods have been discussed in an attempt to approximate it \cite{Igor_paper}. One method is to use geometric discord \cite{geo_discord}, which has found to be bounded by Fisher Information for two qubits \cite{lowe}. However, the benefit of geometric discord is limited as it can increase under local reversible operations \cite{ancilla_problem}. There has been an increased focus in recent years on designing methods for witnessing quantum discord both theoretically, \cite{disc_WITNESS} and even experimentally at the LHC \cite{High_Energy_QD}. Therefore, it is of paramount importance that further techniques are explored in order to witness quantum correlations beyond entanglement, where there is particular focus on devising experimentally feasible protocols.

Following a different approach in order to witness quantum advantage, there has been increasing use of game theory. This is a branch of mathematics which focuses on optimising players decisions in order to achieve their respective best payoffs. Game theory was developed in the middle of the $20^{\text{th}}$ century by Von Neumann and Morgenstern \cite{vonneumann1947} and then extended upon by Nash \cite{Nash_eq}. Typically, game theory allows only classical correlations between the players. However, it is possible to design a game where the players have access to quantum correlations, and thus can use the properties of quantum mechanics in order to achieve a better payoff relative to the respective classical counterpart. This is the origin of quantum game theory. It was pioneered by Meyer \cite{PhysRevLett.82.1052}, and then Eisert, Wilkens, and Lewenstein \cite{1999PhRvL} where the first demonstration of quantum advantage using game-theoretic techniques was proposed. Since then, quantum game theory has been implemented in evolutionary models \cite{Iqbal2000EvolutionarilySS,PhysRevA.65.022306}, and developed onto quantum protocols \cite{LoweIkeda,Ikeda:2021aa} with applications in quantum networks.

Therefore, it is of significant interest to attempt to unite these two fields and subsequently witness quantum advantage using game-theoretic techniques. In particular, it is important to attempt to devise a game which can utilise quantum discord in order to create quantum advantage, as this will be of practical benefit due its robustness against disorder, and therefore it will have real-world application. The type of game that will be studied is a Bayesian game which has only witnessed quantum advantage using a non-local quantum correlation. This was first proposed for the CHSH game \cite{PhysRevLett.23.880} which allows a witness for violation of Bell's inequality \cite{bell}.

In this paper, an extended quantum game is proposed which utilises the properties of quantum discord, in order to witness quantum advantage under a specific set of conditions.
The fundamental difference between classical and quantum correlations is found in the differing probability distributions, thus it is beneficial to introduce probability in quantum mechanics.

In quantum mechanics, the probability is computed using Born's rule, which is defined as
\begin{equation}
P_i = \rm{tr} \Pi_i \rho,
\end{equation}
where $P_i$ is the probability, $\Pi_i$ is the projective measurement, and $\rho$ is the density matrix. For the scenario considered in this paper, the systems considered are bi-partite two particle spin systems, where the density matrix is mixed separable. Given this, the probability to measure the spin of one qubit up or down ($\sigma=\pm 1$) along $\bf a$ and another qubit ($\sigma'=\pm 1$) along ${\bf b}$ (where $\bf a$ and $\bf b$ are Alice and Bob's respective detector settings) is given by;
\begin{equation}
    P(\sigma,\sigma'|{\bf a}, {\bf b})=\rm{tr}\,\Pi_{\sigma|{\bf a}}\otimes\Pi_{\sigma'|{\bf b}}\,\rho\,, 
    \label{prob}
\end{equation}
where
\begin{equation}
\Pi_{\sigma|{\bf a}}=\frac{1}{2}\left[1+\sigma\,{\bf a}\cdot{\bm \sigma}\right]\,,
\end{equation}
where $\boldsymbol{\sigma}$ is the vector of Pauli matrices such that $\boldsymbol{\sigma} = (\sigma_x,\sigma_y,\sigma_z)$. It is important to stress that projective measurements must take place in each subsystem, such that they are separable. If the projector was not separable, then it would entangle the state, and thus any quantum advantage could be due to entanglement. Subsequently, any claims about the quantum correlations being inherently related to quantum discord, would be redundant. Throughout this paper, the measurements take place on the Bloch sphere such that ${\bf{a}} = ( \cos \phi_a \sin \theta_a, \sin\phi_a \sin \theta_a,\cos \theta_a)$ with $|a|=1$, and the assumption that $\phi_a=0$ . Similarly these assumptions are used for $\bf a', b, b'$.

\section{Quantum discord in a modified CHSH game}
The CHSH game is traditionally a Bayesian game, where the players are cooperative. The Bayesian nature arises due to each player having a private prior probability, where the other players are unaware of this. This prior probability is used to compute their expected payoff. For this game, the Nash equilibrium for the game is the maximum payoff that the players can achieve, as any unilateral deviation to their strategy would result in a worse payoff due to the cooperative nature of the game. However, the game has only been studied using either classical states, or non-local states. 
After outlining the relevant aspects of quantum discord, the modified form of the CHSH game is introduced. From this, the modified game is shown to yield quantum advantage, when considering a mixed separable state which has a non-zero quantum discord in a given measurement space.
\subsection{Quantum Discord}
Quantum discord for a given state quantifies quantum correlations by taking the difference between the mutual information between two parties $A$ and $B$ before and after a local projective measurement. This definition reveals how quantum correlations can persist even in the case of zero entanglement. This implies mixed separable states can possess non-zero discord, and therefore non-zero quantum correlation. By defining the mutual information as 
\begin{equation}
I(\rho^{AB}) = S (\rho^A) + S(\rho^B) - S(\rho^{AB}),
\end{equation}
where $S(\rho^{AB}) = - \tr \rho^{AB} \log \rho^{AB} = - \sum_i \lambda_i \ln \lambda_i$, $\rho^{AB}$ is the density matrix of the system, and $\lambda_i$ are the eigenvalues of the density matrix. In order to define the quantum mutual information from player A's perspective given a measurement on player B's subsystem, the sum over all local projective measurements is taken. Therefore, the density matrix after measurement has the form
\begin{equation}
\rho_{A | \Pi_{\mu}^{B}} = \frac{1}{p_{\mu}} \tr_B (\mathbb{1} \otimes \Pi_{\mu}^{B}) \rho^{AB}  (\mathbb{1} \otimes \Pi_{\mu}^{B}),
\end{equation}
where $\Pi_{\mu}^{B} = (1/2)(\mathbb{1}+\mu \bf{n}.\boldsymbol{\sigma})$, $\mu=\pm 1$, where $\bf{n}$ is the Bloch vector. Additionally
\begin{equation}
p_{\mu} = \tr  (\mathbb{1} \otimes \Pi_{\mu}^{B}) \rho^{AB}.
\end{equation}
Note the projector used here only acts in one of the subsystems, not both as defined previously. Therefore quantum discord is defined using local projective measurements in one subsystem. If the quantum discord was computed using $B$'s subsystem given measurements on $A$'s subsystem, then the projective measurements would take place in $A$'s subspace.  From these definitions the conditional entropy can be computed as
\begin{equation}
S(A | \Pi_{\mu}^{B}) = \sum_{\mu} p_{\mu} S(\rho_{A | \Pi_{\mu}^{B}} ).
\end{equation}
Subsequently, the quantum mutual information is
\begin{equation}
J_A (\rho^{AB}) =   S(\rho^A) -  S(A | \Pi_{\mu}^{B}).
\end{equation}
Therefore the quantum discord can be defined as
\begin{equation}
\begin{split}
D_A (\rho^{AB}) &=  \min\limits_{\Pi_{\mu}^{B}} [ I(\rho^{AB}) - J_A (\rho^{AB})] \\&= \min\limits_{\Pi_{\mu}^{B}} S(A | \Pi_{\mu}^{B}) + S(\rho^B) - S(\rho^{AB}).
\end{split}
\end{equation}
Note that due to the minimisation over all projective measurements of player B, any quantum correlations which exist must be due to the initial state, and not due to a particular choice of projector. 
The only states which yield zero quantum discord take the form
\begin{equation}
\rho^{AB} = \rho^A \otimes \rho^B.
\end{equation}
Therefore, generally, states of the form
\begin{equation}
\rho^{AB} = \sum_{\mu=0}^{N} q_{\mu} \rho_{\mu}^{A} \otimes \rho_{\mu}^{B},
\label{genstat}
\end{equation}
result in non-zero discord, where $\sum_{\mu=0}^{N} q_{\mu} =1$. Note how eq. (\ref{genstat}) is clearly a mixed separable state as each subsystem acts in its own space. 
This reveals that quantum correlations can exist beyond entanglement, as will be demonstrated in the solution of the game. 

There are a few extra comments about this definition. It is clear that discord is NP-hard to compute since it is generally a discrete optimisation problem which scales exponentially with the dimensionality of the system. The definition given is the quantum discord for player A. Therefore, the quantum discord for one subsystem is not necessarily the same as for the other subsystem, which implies that quantum discord is in general not symmetric, i.e. $D_A (\rho^{AB}) \neq D_B (\rho^{AB})$. However, to derive the definition for player B follows a similar procedure, but the projective measurements will take place in player A's subspace. Given this definition of quantum discord, the original CHSH game can be introduced to explicitly show how quantum correlations can be revealed and understood using game-theoretic techniques.

\subsection{Description of the game}
The general operational procedure for the CHSH game can be described as follows: a referee, Charlie sends bits either 0 or 1 to Alice, and similarly Charlie sends bits either 0 or 1 to Bob. When Alice receives bit 0, she sets her detector setting to ${\bf{a}}$ and then performs a measurement on her shared state with Bob. When Alice receives bit 1, she sets her detector setting to ${\bf{a'}}$ and performs a measurement on her shared state. When Bob receives bit 0, he sets his detector setting to ${\bf{b}}$, and performs a measurement on their shared state, and when Bob receives bit 1, he sets his detector setting to ${\bf{b'}}$ and performs a measurement. Alice and Bob do not communicate what bits they have received, so they are unaware which game they are playing. Given Alice and Bob's combined measurements, we can assign a payoff. For example, if they both receive bit 0, and therefore are in the type of game given by $({\bf{a}},{\bf{b}})$, and they both measure spin up, then they are assigned a payoff of 1, as shown by the payoff table in Table I. These payoff's are specifically for the CHSH game, therefore any deviation from Table I will result in a modified version of the original game.
\begin{table}[H]
    \centering
     \begin{adjustbox}{width=0.3\textwidth}
    \begin{tabular}{c|c c}
     ($\alpha,\beta$)   & ${\bf{b}}$ &${\bf{b'}}$ \\ \hline
     ${\bf{a}}$  & \begin{tabular}{c|c c}
       &  {$\uparrow$} &{$\downarrow$}\\ \hline
       {$\uparrow$}  & 1 &0  \\
      {$\downarrow$} & 0 & 1
     \end{tabular}  & \begin{tabular}{c|c c}
       & {$\uparrow$} & {$\downarrow$} \\ \hline
       {$\uparrow$} & 1 &0  \\
       {$\downarrow$} & 0 & 1
     \end{tabular} \\ 
     ${\bf{a'}}$  & \begin{tabular}{c|c c}
       &  {$\uparrow$} & {$\downarrow$} \\ \hline
        {$\uparrow$} &  1 &0  \\
        {$\downarrow$} & 0 & 1
      \end{tabular} & \begin{tabular}{c|c c}
       &  {$\uparrow$} & {$\downarrow$} \\ \hline
       {$\uparrow$} & 0  & 1 \\
        {$\downarrow$} & 1 & 0
      \end{tabular}
    \end{tabular}
    \end{adjustbox}
    \caption{This table shows the payoff matrix for the traditional CHSH game. Based on what bits they receive, {\bf{a}},{\bf{b}},{\bf{a'}},{\bf{b}} denote how Alice and Bob's detectors are set up. From this, they then perform measurements on their shared states and based on their results, they are assigned a payoff. Also,  $\alpha=\{{\bf{a}},{\bf{a'}}\}$, $\beta=\{{\bf{b}},{\bf{b'}}\}$ and therefore denotes what bits the players receive, either 0 or 1.}
\end{table}
The general expected payoff to player A for an arbitrary state for this type of game is given by
\begin{equation}
u_A = \sum_{\alpha,\beta,\sigma,\sigma'} u_{A,\sigma,\sigma'}^{\alpha,\beta} P_A(\alpha,\beta) P(\sigma,\sigma' | \alpha,\beta),
\label{exp_pay_A}
\end{equation}
where $u_{A,\sigma,\sigma'}^{\alpha,\beta} $ is a tensor of payoffs for player $A$, $P_A(\alpha,\beta)$ is player A's prior belief of which type of game they will be in, and $P(\sigma,\sigma' | \alpha,\beta)$ represents the conditional probability for measuring a given spin depending on what type of game the players are in. A similar equation written for player $B$'s payoff is given by
\begin{equation}
u_{B} = \sum_{\alpha,\beta,\sigma,\sigma'} u_{B,\sigma,\sigma'}^{\alpha,\beta}  P_B(\alpha,\beta) P(\sigma,\sigma' | \alpha,\beta),
\label{exp_pay_B}
\end{equation}
where $u_{B,\sigma,\sigma'}^{\alpha,\beta}$ are the payoff's for player $B$ and $P_B(\alpha,\beta)$ is player B's prior belief of which type of game they will be in. For the traditional CHSH game, the payoffs assigned to Alice and Bob are the same, as well as each player's prior beliefs, therefore $u_A= u_B$.
Using classical statistics, the best payoff Alice and Bob can achieve is $0.75$. This can be seen if they choose their detector settings such that they will always measure spin up, regardless of the type of game they are in. However, if the players know that they are allowed to share a Bell state, given by
\begin{equation}
\rho = \ket{\psi} \bra{\psi}, \hspace{1cm} \ket{\psi} = \frac{1}{\sqrt{2}} \big( \ket{\uparrow \uparrow} + \ket{\downarrow \downarrow} \big) ,
\end{equation}
then the optimum payoff they can achieve is $(1/4) (2 + \sqrt{2}) \approx 0.85$. In both scenarios, the prior probability is given by $P(\alpha,\beta)=P(\alpha) P(\beta)$ as Alice and Bob are independent, and they each believe $P(\alpha) = P(\beta) = 1/2$, therefore $P(\alpha,\beta)=1/4$. This is justifiable as they have no prior information on what type of game they will be in, so they can only assume each game will be played uniformly. It is important to note that strict quantum advantage (i.e. surpassing 0.75) can only be found in the CHSH game for non-local states. However this does not necessarily mean there is no quantum advantage that can be gained from a mixed separable quantum correlated state. 

\begin{figure}[H]
\centering
\includegraphics[scale=0.58]{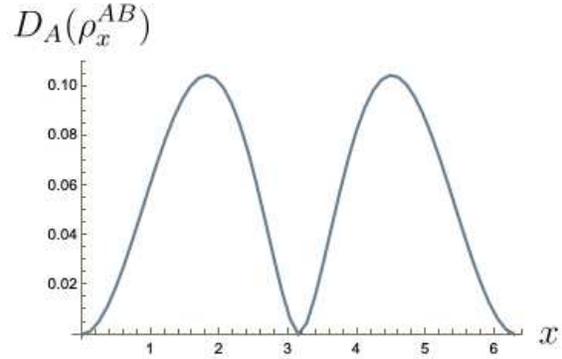}
\caption{This shows the numerical discord for the state in eq. (\ref{diseq}). It is clear there is zero quantum correlation at $x=0,\pi$. Elsewhere, there is non-zero quantum correlation, with the maximum occurring around $x\approx n \pi + \pi/2$, where $n \in \mathbb{Z}$.  The minimisation was over $\theta$, and $\phi$ as defined in the projectors. This can only occur for a 2-qubit state, as the parameterisation for $\theta$ and $\phi$ is only valid for a two particle system. Note $\phi$ is included here for the optimisation to compute the discord, but for the optimisation in the game, $\phi=0$ to allow analytical tractability. }
\label{numDis}
\end{figure}

Consider an extension of the game, where the payoff's are different such that the payoff table is given in Table II.
\begin{table}[H]
    \centering
    \begin{adjustbox}{width=0.35\textwidth}
    \begin{tabular}{c|c c}
      ($\alpha,\beta$)  & ${\bf{b}}$ &${\bf{b'}}$ \\ \hline
     ${\bf{a}}$  & \begin{tabular}{c|c c}
       &  {$\uparrow$} &{$\downarrow$}\\ \hline
       {$\uparrow$}  & -1 & 1  \\
      {$\downarrow$} & 1 & -1
     \end{tabular}  & \begin{tabular}{c|c c}
       & {$\uparrow$} & {$\downarrow$} \\ \hline
       {$\uparrow$} & -1 & 1  \\
       {$\downarrow$} &1 & -1
     \end{tabular} \\ 
     ${\bf{a'}}$  & \begin{tabular}{c|c c}
       &  {$\uparrow$} & {$\downarrow$} \\ \hline
        {$\uparrow$} & -1 & 1 \\
        {$\downarrow$} & 1 & -1
      \end{tabular} & \begin{tabular}{c|c c}
       &  {$\uparrow$} & {$\downarrow$} \\ \hline
       {$\uparrow$} & 1 & -1 \\
        {$\downarrow$} & -1 & 1
      \end{tabular}
    \end{tabular}
    \end{adjustbox}
    \caption{The numbers denote Alice's and Bob's payoff. As before, based on what bits they receive, {\bf{a}},{\bf{b}},{\bf{a'}},{\bf{b'}} denote how Alice and Bob's detectors are set up. From this, they then perform measurements on their shared states and based on their results, they are assigned a payoff.}
\end{table}
This scenario is similar to the traditional CHSH game, where it is a cooperative game and thus both players maximising their payoff's is the Nash equilibrium, as their payoff's are the same.
Unlike the previous CHSH game, a non-entangled state will be used in order to witness quantum advantage utilising quantum correlations. The shared state between Alice and Bob is discorded, where it is defined as
\begin{equation}
\rho^{AB} (x) \equiv \rho_{x}^{AB} = \frac{1}{2} \Big[ \ket{\uparrow \uparrow} \bra{\uparrow \uparrow} + \ket{x x} \bra{x x} \Big],
\label{diseq}
\end{equation}
where $\ket{x} = \cos (x /2) \ket{\uparrow} + \sin (x /2) \ket{\downarrow}$. This is a mixed separable state as it can be written in the same form as eq. (\ref{genstat}), and therefore does not possess any non-locality or entanglement. When numerically computing the discord, the result is given by Fig. (\ref{numDis}).

Since both Alice and Bob know the shared state beforehand, they can tune their respective detector settings in order to maximise their respective payoffs. However, this optimisation includes optimising over each others detector settings. Therefore each others maximum payoff depends on what the other player will do. Fortunately, the cooperative nature of this game yields a solution in which both players would choose the optimum detector settings which would benefit each other.

\subsection{Solution to the game}
At this stage, it is important to emphasise that both Alice and Bob have complete knowledge of each other's payoff tables, and the state which they share. Their only sources of lack of knowledge arise since they do not know what the value of the parameter $x$ will be, and what type of game they will end up playing. Also, it is assumed that the prior $P(\alpha, \beta) = 1/4$ as before. This is because the players have no knowledge of what type of game they will play. Additionally, the state used to compute the conditional probability is given in eq. (\ref{diseq}). Given this knowledge that both Alice and Bob share, they can compute their expected payoff's as functions of each other's choices and the discord parameter $x$. These are given by
\begin{equation}
u_A = u_B = f(\theta_a,\theta_{a'}, \theta_b, \theta_{b'}, x),
\label{sA1}
\end{equation}
where
\begin{equation}
\begin{split}
&f(\theta_a,\theta_{a'}, \theta_b, \theta_{b'}, x) = -\frac{1}{16} \Big[ 2 \cos \theta_{a,b}^{-}  \\ &+ 2 \cos \theta_{a',b}^{-} + 2 \cos \theta_{a,b'}^{-} - 2 \cos \theta_{a',b'}^{-}   + \cos \theta_{a,b}^{+}   \\ &  \cos \theta_{a',b}^{+} + \cos \theta_{a,b'}^{+} - \cos \theta_{a',b'}^{+}  \\ &+ \cos(\theta_{a,b}^{+} - 2x) +  \cos(\theta_{a',b}^{+} - 2x)  \\ &+  \cos( \theta_{a,b'}^{+} - 2x)  -  \cos( \theta_{a',b'}^{+} - 2x)\Big],
\label{f}
\end{split}
\end{equation}
where $\theta_{\alpha,\beta}^{-} = \theta_\alpha - \theta_\beta$ and $\theta_{\alpha,\beta}^{+} = \theta_\alpha + \theta_\beta$. This expected payoff can be re-written by adding and subtracting $\cos \theta_{a,b}^{+} +\cos \theta_{a,b'}^{+} +\cos \theta_{a',b}^{+} -\cos \theta_{a',b'}^{+}$ such that eq. (\ref{f}) becomes
\begin{equation}
\begin{split}
&f(\theta_a,\theta_{a'}, \theta_b, \theta_{b'}, x) = -\frac{1}{16} \Big\{ 2 \Big[ \cos \theta_{a,b}^{-}  \\ &+  \cos \theta_{a',b}^{-} +  \cos \theta_{a,b'}^{-} -  \cos \theta_{a',b'}^{-}   + \cos \theta_{a,b}^{+}   \\ &  \cos \theta_{a',b}^{+} + \cos \theta_{a,b'}^{+} - \cos \theta_{a',b'}^{+} \Big] \\ &+ \cos(\theta_{a,b}^{+} - 2x) +  \cos(\theta_{a',b}^{+} - 2x)  \\ &+  \cos( \theta_{a,b'}^{+} - 2x)  -  \cos( \theta_{a',b'}^{+} - 2x) \\& - \cos \theta_{a,b}^{+} -\cos \theta_{a,b'}^{+} -\cos \theta_{a',b}^{+} +\cos \theta_{a',b'}^{+}\Big\},
\label{f1}
\end{split}
\end{equation}
It is important to emphasise that this has not changed the result, but now allows eq. (\ref{f1}) to be written into a purely classical term, and a purely quantum term. 
Finally, the expected payoff can be written as
\begin{equation}
\begin{split}
f(\theta_a,\theta_{a'}, \theta_b, \theta_{b'}, x) &= f_{Cl} (\theta_a,\theta_{a'}, \theta_b, \theta_{b'}) \\&+ f_{Q}( \theta_a,\theta_{a'}, \theta_b, \theta_{b'}, x),
\end{split}
\label{fTot}
\end{equation}
where the classical term is given by
\begin{equation}
\begin{split}
& f_{Cl} (\theta_a,\theta_{a'}, \theta_b, \theta_{b'}) = - \frac{1}{4} \Big[ \cos \theta_a ( \cos \theta_b \\&+ \cos \theta_{b'} ) + \cos \theta_{a'} ( \cos \theta_b - \cos \theta_{b'} ) \Big],
 \end{split}
 \label{fCl}
 \end{equation}
and the quantum term has the form
\begin{equation}
\begin{split}
& f_{Q} (\theta_a,\theta_{a'}, \theta_b, \theta_{b'},x) = - \frac{\sin x}{8} \Big[  \sin(\theta_{a,b}^{+} - x) \\&+  \sin(\theta_{a',b}^{+} - x)  +  \sin( \theta_{a,b'}^{+} - x)  \\&-  \sin( \theta_{a',b'}^{+} - x) \Big].
\end{split}
\label{fQ}
\end{equation}
Due to the $\sin x$ term, eq. (\ref{fQ}) becomes zero at $x=0,\pi$ and non-zero elsewhere. This is a truly quantum contribution, which can be seen by noting that the quantum correlations for the shared state vanish when $x=0,\pi$, seen explicitly in Fig. \ref{numDis}. 

\begin{figure}[H]
\centering
\includegraphics[scale=0.4]{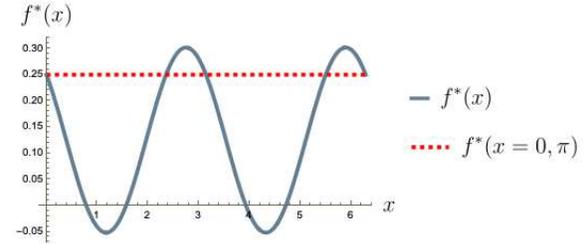}
\caption{This plot reveals how there are regions where there is quantum advantage, for given different values of $x$, such that the expected payoff is greater than $0.25$. The maxima of this function are found at $7\pi/8$ and $15\pi/8$, which yields 0.30178.  }
\label{fstar}
\end{figure}

Given it is possible to split the expected payoff into two contributions, coming from classical correlations and quantum correlations, opens up the possibility that there may be some advantage to be gained from the extra quantum term. By maximising over all the variables with no restrictions, the maximum expected payoff is $0.5$ which coincides with the largest classical payoff. This is expected, as the state considered is a mixed separable state, and surpassing this bound would imply non-locality. However, by restricting the measurement space available to the players, such that they are only allowed to set their measurement angles from 0 to $\pi/2$, yields an unexpected quantum advantage for the players. Numerically maximising eq. (\ref{fTot}), it is found that 
\begin{equation}
f^{max} (\theta_{a}^{*},\theta_{a'}^{*}, \theta_{b}^{*}, \theta_{b'}^{*}, x^{*}) = 0.30178 \equiv f^{*},
\end{equation}
where $\theta_{a}^{*} = \theta_{b}^{*} = \pi/2$, $\theta_{a'}^{*} = \theta_{b'}^{*}=0$, and crucially $x^{*}=7 \pi / 8$. The key result is that a non-zero $x$ gave the maximum for the expected payoff. For comparison, by considering a purely classical scenario ($x=0$), the maximum expected payoff is 0.25. This is emphasised in Fig. \ref{fstar}.
It is also possible to consider scenarios where the quantum correlation dominates the classical correlation. Defining the ratio between the quantum contribution and the classical contribution, yields
\begin{equation}
\begin{split}
& \kappa(\theta_a,\theta_{a'}, \theta_b, \theta_{b'},x) = \abs{\frac{f_Q(\theta_a,\theta_{a'}, \theta_b, \theta_{b'},x) }{f_{Cl} (\theta_a,\theta_{a'}, \theta_b, \theta_{b'})}} \\ &\equiv \kappa,
\end{split}
\end{equation}
where the quantum contribution must obey $f_Q(\theta_a,\theta_{a'}, \theta_b, \theta_{b'},x) \geq 0$ and similarly the classical part is constrained such that $f_{Cl} (\theta_a,\theta_{a'}, \theta_b, \theta_{b'}) \geq 0$. Similar ratios could be defined for differing signs.
Therefore, when $\kappa > 1$, this implies the quantum contribution is larger than the classical contribution, and conversely $\kappa < 1$ implies the classical contribution is dominating the quantum contribution. When the contributions from both the classical and quantum part are equal, then $\kappa = 1$. This ratio gives a measure of how much quantum correlation is in a given quantum game relative to the classical counterpart. It is important to emphasise, the total correlation may not reveal quantum advantage in the particular quantum game. However, it is possible that the classical correlation is zero, and there is a non-zero quantum correlation resulting in an infinite $\kappa$. This is the case for $\theta_a = \pi/2$, $\theta_{a'} = \pi/4$, and $\theta_b = \theta_{b'} = 0$. Whether it is possible to develop a hierarchy between classical and quantum correlation below the CHSH bound remains an open question. 

\subsection{Relations}
There are interesting relations which can be derived from the results of the game. These can be analysed by considering eqns. (\ref{fCl}) and (\ref{fQ}) individually, which are given by
\begin{equation}
\begin{split}
&f_{Cl}^{max}  (\theta_{a}^{*}=\pi/2 ,\theta_{a'}^{*} = 0, \theta_{b}^{*} = \pi/2, \theta_{b'}=0) \\ &= 0.25 = \frac{4}{16},
\end{split}
\end{equation}  
and 
\begin{equation}
\begin{split}
& f_{Q}^{max}  (\theta_{a}^{*}=\pi/2 ,\theta_{a'}^{*} = 0, \theta_{b}^{*} = \pi/2, \theta_{b'}=0, \\&x=7 \pi /8) \approx 0.05178 = \frac{0.828}{16},
\end{split}
\end{equation}
which interestingly when combined can be written as
\begin{equation}
\begin{split}
f^{max} (\theta_{a}^{*},\theta_{a'}^{*}, \theta_{b}^{*}, \theta_{b'}^{*}, x^{*})&\approx \frac{1}{16} \big( 4 + 0.828 \big) \\&= \frac{1}{16} \big( 2 + 2\sqrt{2} \big).
\end{split}
\label{QgamMax}
\end{equation}
This can be directly compared with the result from the CHSH game using a non-local state, where the differences are due to the denominator, and an extra factor of 2 in front of $\sqrt{2}$.
The apparent connection to the maximum bound for the CHSH inequality of $2 \sqrt{2}$ was not expected as the state considered for maximum violation is strictly non-local. Despite this, it is clear that the extra quantum term in the game considered here gives rise to an extra contribution which can be utilised for quantum advantage in this Bayesian game.

When calculating the maximum expected payoff, it was noted that an interesting relation occurs between this expected payoff and the trace of the Hessian matrix. Given the trigonometric nature of the expected payoff, it is clear that the second derivatives will return the same functions with the signs changed. The sum of the second derivatives with respect to their corresponding variables is proportional to the expected payoff such that 
\begin{equation}
\tr \hat{H} = - 2 f(\theta_a,\theta_{a'}, \theta_b, \theta_{b'}, x),
\end{equation}
where $\hat{H}$ is the Hessian matrix. Since
\begin{equation}
\tr \hat{H} = \sum_i \phi_i,
\end{equation}
where $\phi_i$ are the eigenvalues of the Hessian matrix yields the relation that
\begin{equation}
 f(\theta_a,\theta_{a'}, \theta_b, \theta_{b'}, x)=- \frac{1}{2} \sum_i \phi_i.
\end{equation}
Therefore, optimising over the eigenvalues of the Hessian matrix is equivalent to optimising over the expected payoff function. It is important to stress, this has only been found for this particular payoff function, but whether there is an underlying relation between the eigenvalues of the Hessian and the expected payoff in Bayesian game theory remains an interesting direction for future research.

\section{Discussion}
The proposed game offers tangible benefit for experimentalists when building quantum discorded states. Given the algorithmic procedure of this game, this operational method gives a framework in which the experimentalists could verify that the state they have attempted to build is correct. For example, an experimentalist would know the state they have designed gives quantum advantage when performing the proposed analysis of this paper using the given constraints. This work could be extended such that, when attempting to build other states, similar calculations can be performed and further quantum advantage could be witnessed.

There are two significant limitations of the analysis performed in this paper. The main drawback is that quantum advantage was demonstrated for a specific discorded state, rather than proving quantum advantage for a general discorded state. However, this paper opens up the avenue for developing methodology which can witness quantum advantage using a general discorded state in a quantum game.  Subsequently, this will be a key focus of future research.

Another significant limitation, is due to the restrictions imposed on Alice and Bob's measurements. Whilst it is beneficial for witnessing quantum correlations, the maximum classical expected payoff will always be the upper bound for mixed separable states. To further generalise the results, $\phi_{a}$ would also be generally non-zero, similarly for ${\bf{ a'}} ,{\bf{ b}}$ and ${\bf{ a'}}$. Whether this allows further quantum advantage would be a significant question to investigate. 

It would be interesting to determine if there is a generic relation between the measurement space and the ability to witness quantum advantage. The same logic could be applied; if given a specific quantum state, is there a measurement space which would witness the quantum correlation. 

From a game-theoretic perspective, it would be interesting to modify this game such that Alice and Bob are competing, thus no longer cooperative. Given this change, how would the quantum correlations affect the Nash equilibrium compared to the classical solution. This may offer further insight into the benefit of utilising quantum correlations in game theory. As a further extension of this, this quantum game could be implemented onto a quantum network, where a quantum protocol could be devised which could offer technological benefit by utilising quantum technologies. It is also worth studying whether quantum advantage beyond entanglement can be witnessed in extensive form games, as quantum advantage has already been shown for non-local states \cite{Ikeda:2023aa}.

It was also interesting to note the similarities between the maximum expected payoff in the CHSH game, and the result in eq. (\ref{QgamMax}). Given the discorded state is inherently local, it is intriguing to see a factor of $2 \sqrt{2}$ contributing to the quantum advantage.  A further unexpected result that arose from the analysis was the relation between the spectra of the eigenvalues of the Hessian matrix, and the expected payoff function. Further investigation is required to determine whether this is a general relation.

\section{Conclusion}
Quantum advantage is witnessed in a modified version of the CHSH game, where the advantage is entirely due to quantum correlations beyond entanglement. This proof of concept result was done by taking a specific discorded state, and imposing measurement restrictions on the players. Using these constraints, the players were able to surpass the classical limit for the given restrictions, entirely due to the quantum correlations. By utilising game-theoretic techniques, this protocol could be readily implemented in experiments, in order to witness and characterise quantum advantage.

This is the first theoretical result using quantum discord for quantum advantage in Bayesian game theory. Future research arising from this work will extend quantum discord into a wide range of quantum games, in order to witness and quantify quantum advantage, with specific focus on the importance of the measurement space.

\section{Acknowledgements}

The author thanks David Lowe for insightful discussions about incomplete information games, Igor Yurkevich for assistance with probability in quantum mechanics and Kazuki Ikeda for discussing the applications of quantum game theory. This work was supported by the Leverhulme Trust Grant
No. RPG-2019-317.

\bibliographystyle{apsrev4-1} % Tell bibtex which bibliography style to use

%\bibliography{Literature_Review_References}


\end{document}